\documentclass[11pt]{article}
\usepackage{graphicx}
\usepackage[margin=1.25in]{geometry}
\usepackage[usenames,dvipsnames]{color}
\usepackage{float}
\usepackage{amssymb}
\usepackage{amsmath}
\usepackage{slashed}
\usepackage{array}
\usepackage{lineno}
\usepackage{scrextend}
\usepackage{url}
\usepackage[colorlinks = true,
            linkcolor = blue,
            urlcolor  = blue,
            citecolor = blue,
            anchorcolor = blue]{hyperref}

\newcommand\pubnumber{}
\newcommand\pubdate{\today}

\def\Title#1{\begin{center} {\LARGE #1 } \end{center}}
\def\Author#1{\begin{center}{ \sc #1} \end{center}}

\newcommand\pubblock{\rightline{\begin{tabular}{l} \pubnumber\\
         \pubdate \end{tabular}}}
\newenvironment{Abstract}{\begin{quotation} \begin{center}
                       ABSTRACT
     \end{center}\bigskip  }{\end{quotation}}

\input Common/workshopsymbols.tex

\newcommand\snowmass{\begin{center}\rule[-0.2in]{\hsize}{0.01in}\\\rule{\hsize}{0.01in}\\
\vskip 0.1in Submitted to the  Proceedings of the US Community Study\\ 
on the Future of Particle Physics (Snowmass 2021)\\ 
\rule{\hsize}{0.01in}\\\rule[+0.2in]{\hsize}{0.01in} \end{center}}

\def\meg{\mu^+ \rightarrow e^+ \gamma}
\def\menng{\mu^+ \rightarrow e^+ \nu_e \bar{\nu_\mu} \gamma}
\def\meee{\mu^+ \rightarrow e^+ e^- e^+}
\def\mec{\mu^-N\rightarrow e^-N}
\def\meX{\mu^+ \rightarrow e^+ X}
\def\megX{\mu^+ \rightarrow e^+ \gamma X}

\begin{document}

\pubblock
\Title{\bf Charged Lepton Flavor Violation \\
\small \vspace{0.2cm} Report of Topical Group RF5\\
Rare Processes and Precision Measurements Frontier}
\bigskip 

\Author{Conveners: S.~Davidson, B.~Echenard}
\Author{Contributors: R.H.~Bernstein, J.~Heeck, D.G.~Hitlin }

\begin{Abstract}
\noindent This reports summarizes the activities of the Charged Lepton Flavor Violation (CLFV) group of the 2022 Community Summer Study. CLFV reactions provide unique information on the scale and dynamics of flavor generation, and more generally a wide range of New Physics scenarios, complementing direct searches performed at collider and neutrino physics experiments. These processes already probe mass scales up to thousands of TeV, and an observation would be an unambiguous signature of physics beyond the Standard Model. We review the current status and future experimental opportunities in muon, tau and heavy state transitions, with a focus on US-led initiatives. 
\end{Abstract}

\snowmass

\def\thefootnote{\fnsymbol{footnote}}
\setcounter{footnote}{0}
\newpage
\section{Executive summary}

The broad physics potential of charged lepton flavor violating (CFLV) reactions, and the discovery opportunities at related experiments, have been much emphasized during the Snowmass process. CLFV processes open a unique window on many New Physics (NP) scenarios, complementing searches performed at collider, dark matter, and neutrino physics experiments. Existing studies provides crucial information on the scale and dynamics of flavor generation, probing mass scales at the level of $10^3 - 10^4 \TeV$. A CLFV observation would be unambiguous evidence of new particles beyond the Standard Model. 

Impressive sensitivity gains are expected in the near future, with up to four orders of magnitude improvements in the rate of $\mec$ conversion and $\meee$ decay searches, and up to two orders of magnitude in $\tau$ CLFV decays. The expected ratio among the  (relative) rates for  these processes depends on the NP amplitude-generating flavor violating effects, and comparisons offer powerful model discrimination. In the case of muons, the relative strength among the decay and conversion processes is driven by a few parameters governing the contribution of dipole-type and four-fermion operators. As an example, the reach and complementarity of current and proposed experiments derived in Effective Field Theory is illustrated in Figure~\ref{fig:NPreach}. Existing measurements already place strong constraints on the NP mass scale probed by these operators, and the planned improvements will significantly extend our knowledge in the coming decade. On the other hand, tau CLFV searches can be conducted over many final states, which is promising for identifying the nature of the underlying NP.
Decays of heavy states ($Z$,$h$,...) and mesons would give another handle on the structure of physics beyond the SM. 

A global experimental program of CLFV searches is underway in the US, Europe and Asia. Among the most sensitive probes are experiments using high intensity muon beams to search for CLFV transitions: $\meg$ and $\meee$ decays with MEG-II and Mu3e at PSI, and the coherent neutrinoless conversion of a muon into an electron, $\mec$, with Mu2e at FNAL and COMET at J-PARC. Upgrades to the beamlines at PSI, Fermilab, and J-PARC offer the possibility to further extend the discovery potential. In particular, a staged program of next-generation experiments and facilities has been proposed at FNAL to exploit the full potential of the PIP-II accelerator. Mu2e-II is a near-term evolution of the Mu2e experiment, with an order of magnitude or more improvement in sensitivity to the conversion rate. By leveraging existing infrastructures, Mu2e-II plans to starts construction before the end of the decade. The Advanced Muon Facility is a more ambitious proposal for a new high-intensity muon complex in the next decade. This facility would provide the world's most intense positive and negative muon beams, enabling a suite of experiments with unprecedented sensitivity to probe mass scales in the range $10^4 - 10^5 \TeV$, as well as the possibility to identify the type of operators contributing to New Physics. The development of this complex has also synergies with R\&D for the muon collider and a beam dump dark matter program at FNAL. In tau decays, Belle-II at SuperKEKB promises great improvement in sensitivity over many channels, and the addition of polarized electron beams could provide additional gains. Complementing these efforts, the HL-LHC and the next-generation of high-energy colliders will continue to explore heavy state decays, while Belle-II and LHCb will play a leading role in searching for CLFV in meson decays.


CLFV searches confront the lepton sector in unique way, and may provide the next clues to understanding physics beyond the Standard Model. The next generation of CLFV experiments and facilities are an essential component of a global effort to search for NP. In particular, a staged program comprising a near term upgrade of the Mu2e experiment, followed by a new high-intensity muon complex at FNAL in the next decade, would offer unique possibilities to study CLFV reactions with unprecedented sensitivities. Strong and continued support of the US community towards R\&D and the realization of this program is critical to fully understand many aspects  of physics beyond the Standard Model. 
\section{Introduction}

The observation of neutrino oscillations provided evidence that charged lepton must experience lepton-flavor-violating contact interactions~\cite{Gonzalez-Garcia:2002bkq}. If neutrino masses arise similarly to those of other fermions (via Yukawa interactions with the SM Higgs), CLFV rates are  suppressed in the SM to unobservably small levels~\cite{Petcov:1976ff, Marciano:1977wx, Lee:1977qz, Lee:1977tib, Hernandez-Tome:2018fbq, Blackstone:2019njl}. For example, the $\mu \rightarrow e \gamma$ decays branching fraction is of the order of $10^{-54}$, well below the sensitivity of any practical experiment. However, neutrino masses could be generated via a different mechanism (see e.g.~\cite{Gonzalez-Garcia:2002bkq, Gehrlein:2022nss, Arguelles:2022xxa, Almumin:2022rml}), giving rise to potentially large CLFV effects. More generally, many scenarios of physics beyond the Standard Model (SM) introduce new sources of CLFV~\cite{Altmannshofer:2022aml}, leading to rates that may be accessible at the next generation of experiments. An observation would be an unambiguous sign of NP. It might also shed light on the neutrino mass mechanism, and even the matter excess of the Universe, should it arise from  leptogenesis~\cite{Fukugita:1986hr,Davidson:2008bu,Chun:2017spz}.

Where to look for CLFV? This question has frequently been  addressed by theorists; there are motivated arguments in favor of various channels, and a variety of predictions from models~\cite{Calibbi:2017uvl} --- some models even predict both the mass scale and flavor pattern of the New Physics~\cite{Vicente:2014wga, Heeck:2018ntp}. The new particles are often  assumed to be heavy, leaving contact interactions among SM particles as low-energy footprints as discussed in the next Section. But the new particles could also be light, for which dedicated searches are planned~\cite{Hesketh:2022wgw, CGroup:2022tli, Mu2e-II:2022blh}. In the absence of strong guidance, the upcoming experimental program will search under all lampposts:

\begin{itemize}
\item Muon transitions have already produced some of the best constraints on CLFV processes; the next generation of experiments will further improve these bounds. In particular, $\mu \rightarrow e$ experiments aim to probe branching ratios four orders of magnitude beyond the current limits~\cite{CGroup:2022tli}, reaching an impressive sensitivity to NP scale beyond $10^4 - 10^5 \TeV$. The conversion rate (on nuclei) as a function of the target material can also provide information about the NP structure. The discrimination power depends on the specific targets, although a combination of low-$Z$ and high-$Z$ materials usually offer good complementarity.

\item The $\tau \rightarrow l$ sector is promising for identifying the underlying NP, due to the comparable sensitivity to a multitude of observables. Flavor changing rates involving the third generation are larger in many CLFV scenarios and in the quark sector of the SM. A variety of next-generation experiments will improve the sensitivity by one to two orders of magnitude and probe scales $\Lambda_{NP}\gsim 10 \TeV$~\cite{Banerjee:2022xuw}.

\item Processes changing lepton flavor or lepton number by two units are particularly interesting since they are linked to Majorana neutrino masses through the Black Box theorem~\cite{PhysRevD.25.2951}. Examples of $\Delta L=2$ processes include muonium-antimuonium oscillations~\cite{Conlin:2020veq, Fukuyama:2021iyw} and a few  meson  decays such as $K^+ \to \pi^-\mu^+e^+$ \cite{Littenberg:2000fg}.

\item Collider experiments directly probe CLFV interactions of heavy particles, that  only contribute indirectly to low-energy processes. These reactions provide complementary constraints, which are comparable to those obtained at lower energies~\cite{HeavyStates}.

\item Meson decays provide an additional window on CLFV interactions, with a unique sensitivity to NP models that connect lepton and quark flavor change (as arises for example in leptoquark models favored by current $B$ anomalies~\cite{Cornella:2019hct}). This topic briefly summarized in this report, a more comprehensively discussion can be found in the "Weak decays of b and c quarks" and "Weak decays of strange and light quarks" Topical Group reports~\cite{rapportTG1,rapportTG2}.

\item Next generation experiments searching for CLFV lepton decays could investigate light NP with CLFV couplings in decays~\cite{LoI080,Echenard:2014lma, Calibbi:2020jvd}, such as $\mu \to e X$ or $\mu\to e\gamma X$, and probe flavor-diagonal feeble couplings in $ \mu\to e\nu \bar{\nu} X$ reactions~\cite{Hesketh:2022wgw}.     

\item Finally, some NP models (strongly) correlate LFV and lepton flavor conserving observables, such as electric dipole moments and anomalous magnetic moments. These observables are discussed in the "Fundamental Physics in Small Experiments" Topical Group report~\cite{rapportTG3}.

\end{itemize}

This report will discuss theoretical aspects of CLFV reactions, and review current experimental efforts and future initiatives in muon, tau, meson, and heavy particle transitions. Some emphasis is given on the muon sector as there are tantalizing suggestions from $B$ decays~\cite{LHCb:2022ine} and the anomalous magnetic moment of the muon~\cite{Muong-2:2021ojo, Aoyama:2020ynm} that NP might be almost within reach.

\section{Theory}

\subsection{Effective Field Theory}
New particles responsible for CLFV are often taken to be heavy, allowing the resulting lower energy predictions to be parameterized using Effective Field Theory (EFT), as summarized in this section. A wide selection of models are  described in~\cite{Calibbi:2017uvl}. In the case of light CLFV new particles, specific models~\cite{Calibbi:2020jvd} and searches~\cite{Hesketh:2022wgw, CGroup:2022tli, Mu2e-II:2022blh} are required. 

Effective Field Theory is a convenient theoretical framework in which to assess the impact of CLFV searches across various energy scales. For instance, the CLFV muon decays $\meee$ and $\meg$, and (Spin Independent) $\mec$ conversion can be parameterized by the following Lagrangian at the experimental scale $\sim m_\mu$ 
\begin{align}
&\mathcal{L}_{\mu e} = -\frac{4 G_F}{\sqrt{2}} \sum_{X = L,R}\left[ m_\mu C_{D,X}\, \overline{e} \sigma^{\alpha\beta} P_X \mu \,F_{\alpha\beta} + C_{S,XX}\, \overline{e} P_X \mu\, \overline{e} P_X e \right. \label{eq:lagr} \\
 &\left. +
 \sum_{Y\in L,R} C_{V,XY} \,\overline{e}\gamma^\alpha P_X \mu \, \overline{e} \gamma_\alpha P_Y e\
 +\sum_{N=p,n}\left( C_{S,X}^{N}\, \overline{e} P_X \mu \, \overline{N} N
+ C_{V,X}^{N}\, \overline{e} \gamma^\alpha P_X \mu \, \overline{N} \gamma_\alpha N \right)\right]
\nonumber
\end{align}
where $P_{L,R}$ are chiral projection operators, and Spin Dependent conversion~\cite{Cirigliano:2017azj, Davidson:2017nrp, Hoferichter:2022mna} is neglected because it occurs at a relatively suppressed rate. The $C_a$ are dimensionless Wilson coefficients, which can be calculated in terms of model parameters when the underlying model is known. As long as the scale of NP is much greater than the $\GeV$ scale, the above Lagrangian provides a model-independent description of CLFV interactions involving muons, electrons, and nucleons, at leading order in $\chi$PT. A similar Lagrangian, but with more operators, could describe $\tau$ decays. 

\begin{figure}[htb]
\begin{center}
 \includegraphics[width=0.45\textwidth]{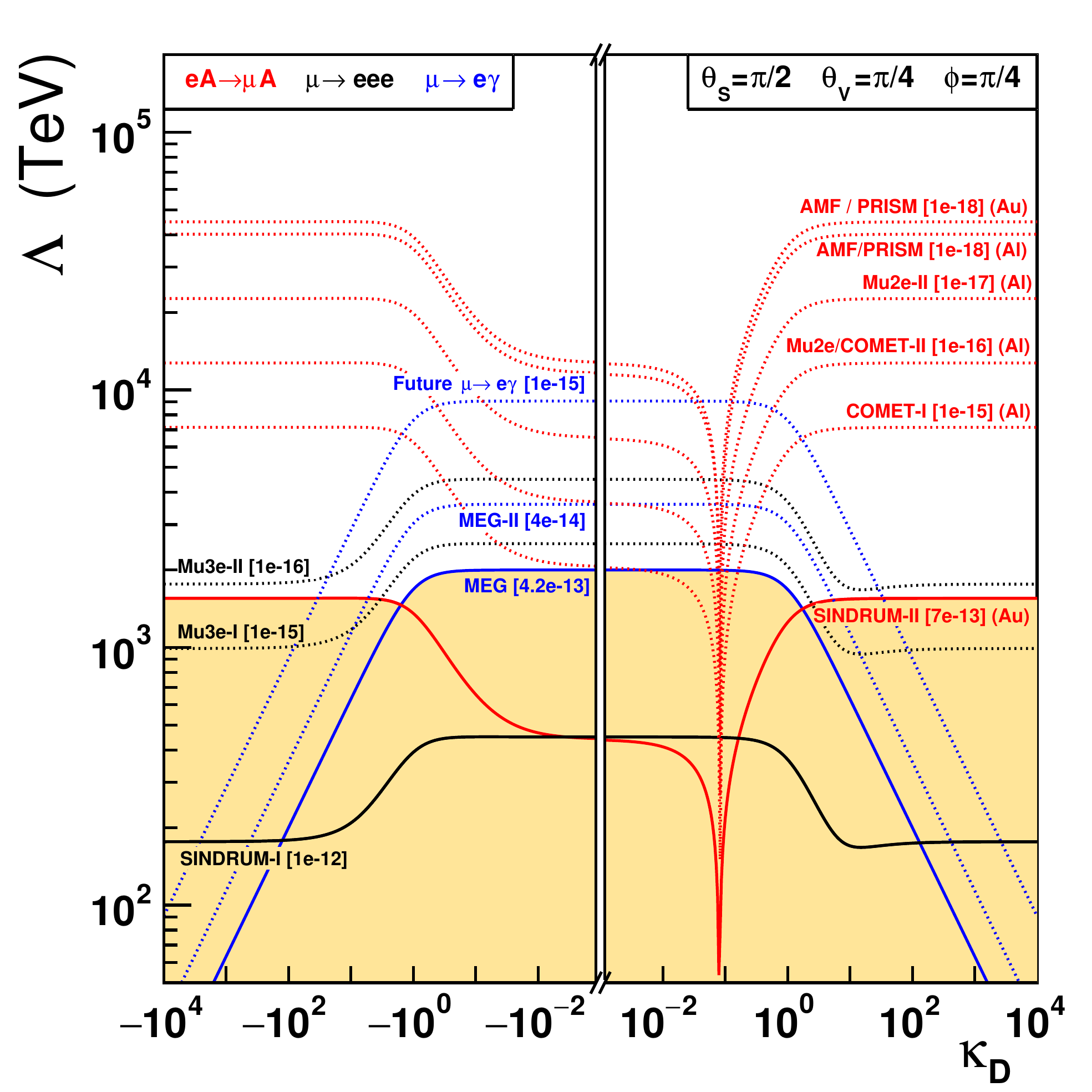}
 \includegraphics[width=0.45\textwidth]{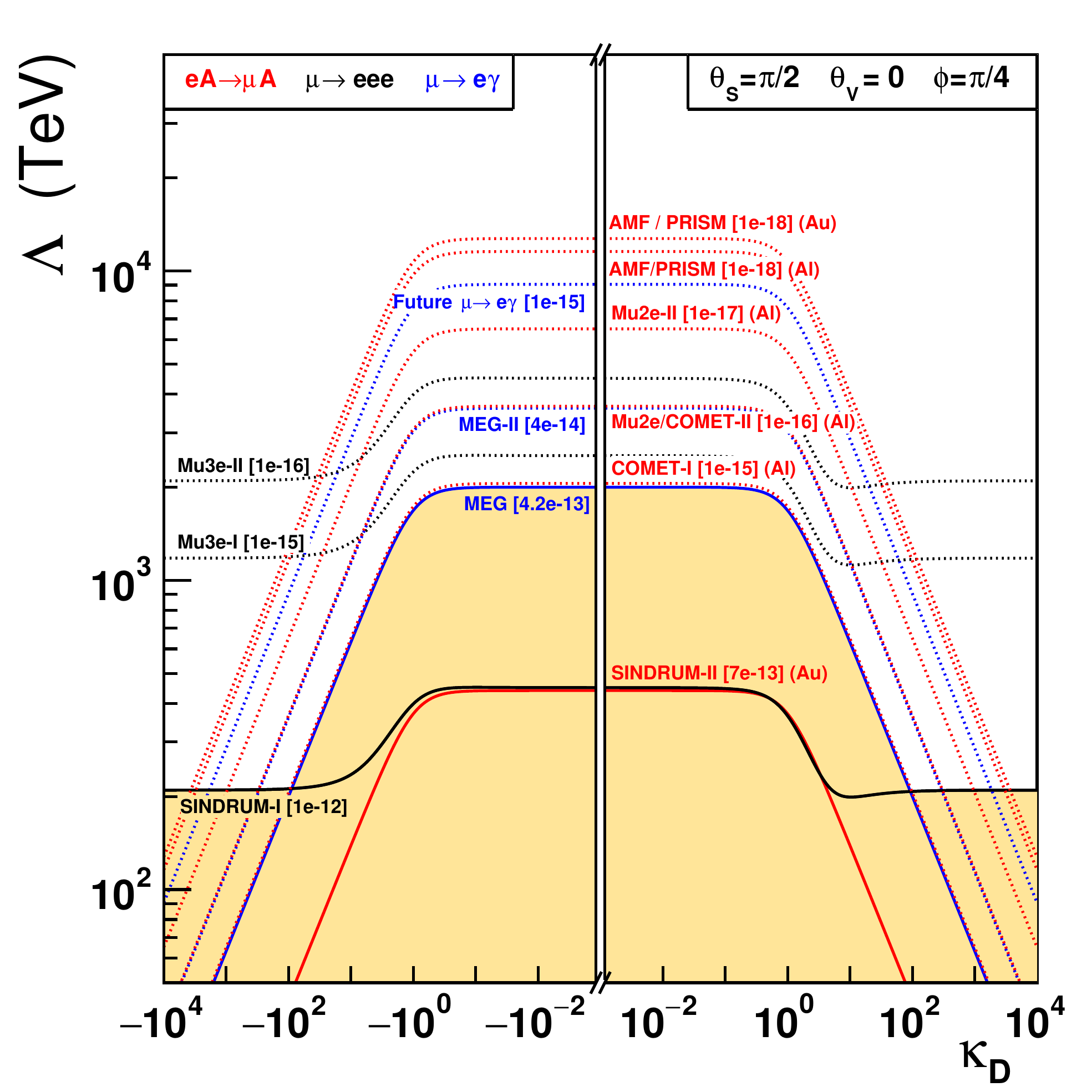}
\end{center}
\caption{Reach in NP scale, $\Lambda$, of past and upcoming $\mu \to e$ searches. The solid region is currently excluded. The parameter $\kappa_D$ is related to the relative contribution of the dipole and the four-fermion contact operators. For $|\kappa_D| \ll 1$ the dipole operator dominates, while the four-fermion operators dominate for $|\kappa_D| \gg 1$. 
The remaining parameters are fixed to typical values (see~\cite{Davidson:2022nnl} for details).
\label{fig:NPreach} }
\end{figure}

The reach and complementarity of past and upcoming $\mu\to e$ experiments is illustrated in Fig.~\ref{fig:NPreach} (taken from~\cite{Davidson:2022nnl}). The curves are obtained by translating the coefficients of Eq.~(\ref{eq:lagr}) to the NP scale via Renormalisation Group Equations (RGEs), and parameterizing them in spherical coordinates (so $2\sqrt{2} C_D = |e_D|\cos(\theta_D)/\Lambda^2$, where $|e_D|\sim 1$ encodes RGE effects). The variable $\kappa_D = {\rm cotan}(\theta_D)-\pi/2$\footnote{The constant $-\pi/2$ is added to be consistent with the definition of the variable $\kappa$ in ~\cite{deGouvea:2013zba}.} describes the relative contribution of the dipole and selected four-fermion  operators, with logarithmic measure:  for $|\kappa_D| \ll 1$ the dipole operator dominates, while the four-fermion operators dominate for $|\kappa_D| \gg 1$. The variable $\theta_V$  denotes the angle between four-fermion operators on leptons or quarks, and $\phi$ distinguishes coefficients that can be probed by $\mu-e$ conversion on Al  vs. Au.

\begin{figure} [htb]
 \centering
 \includegraphics[width=3in]{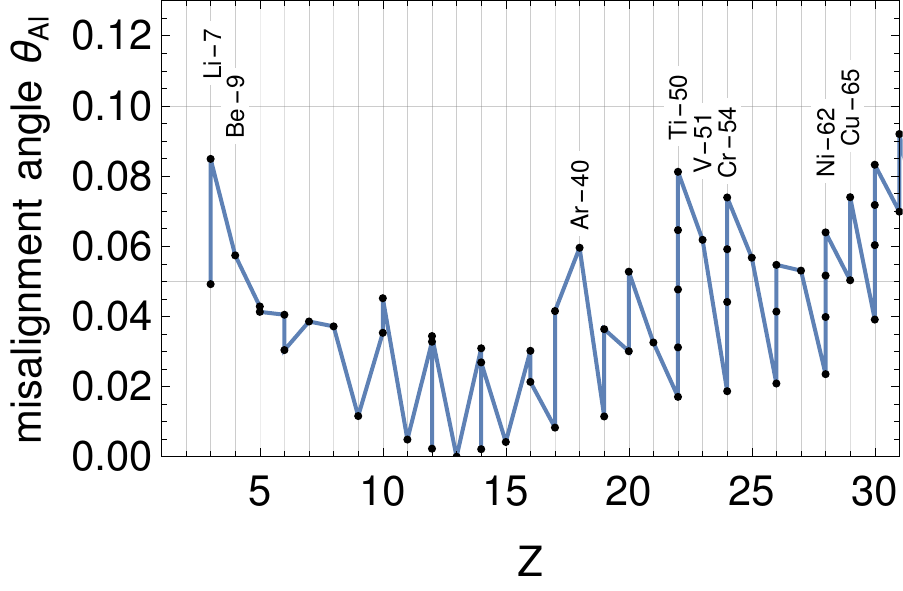}
 \caption{Misalignment angle with Al, taken from~\cite{Heeck:2022wer}. The misalignment angle increases with the number of neutrons in isotopes.}
 \label{fig:Al_misalignment_isotopes_zoom}
\end{figure}

When $\mu\to e \bar{e}e$ is observed, the dipole and four-lepton operators can be distinguished in the final state angular distributions with polarized muons~\cite{Okada:1999zk}. In Spin-Independent $\mu\to e$ conversion, all operators add coherently at the amplitude level, weighted by nucleus-dependent overlap integrals~\cite{Kitano:2002mt,Heeck:2022wer}. As a result, changing the nuclear target probes a different combination of operators~\cite{Kitano:2002mt,Davidson:2018kud}. If the overlap integrals for each nucleus are represented as a vector in the space of operator coefficients, then the complementarity of different nuclei can be represented as the misalignment angle between different vectors~\cite{Mu2e-II:2022blh, Davidson:2018kud}. As pointed out long ago, light and heavy targets provide good complementarity, so an ideal second target after Al would be heavy --- say Au or Pb. Within the Mu2e-II~\cite{Mu2e-II:2022blh} and COMET experiments, this is not possible due to the short muon lifetime in heavier elements~\cite{MuonLifetime}. Focusing on targets with $Z<25$, Fig.~\ref{fig:Al_misalignment_isotopes_zoom} shows that Li-7 and Ti-50 have larger complementarity with respect to Al~\cite{Heeck:2022wer}. This could ultimately help to distinguish CLFV operators involving protons from those involving neutrons.

EFT not only gives a generic parameterization of NP effects, as illustrated in Eqn (\ref{eq:lagr}), it also elegantly separates known SM dynamics from the unknown NP (in the operator coefficients). This means that the nucleon operators can be matched onto operators involving quarks (see {\it e.g.}~\cite{Davidson:2022nnl} for references), and SM loop corrections can be included through the RGEs for the operator coefficients. The QED and QCD loops below the weak scale are relevant for the $\mu\to e$ sector~\cite{Crivellin:2017rmk, Davidson:2020hkf}, because they ensure that almost all the $\mu\to e$ operators with four or fewer legs will contribute to $\mu\to e \gamma$, $\mu\to e\bar{e} e$ and/or $\mu\to e$ conversion, suppressed at most by a factor $\gsim 10^{-3}$.

Beyond the weak scale $v = 1/(\sqrt{2} G_\text{F})^{1/2} \simeq 246$~GeV, heavy New Particles (that are weakly coupled\footnote{An alternative Higgs EFT (HEFT)\cite{Buchalla:2013rka,Pich:2018ltt}, can be appropriate when there are contributions to electroweak symmetry breaking beyond the SM Higgs.}) can be described by the Standard Model EFT (SMEFT)~\cite{Weinberg:1979sa,Wilczek:1979hc,Buchmuller:1985jz,Grzadkowski:2010es,Jenkins:2013zja,Jenkins:2013wua,Alonso:2013hga}, where their effects are encoded in operators of dimension greater than four built out of SM fields that are suppressed by inverse powers of the ``New Physics'' scale $\Lambda_{NP}$ 
\begin{equation}
 {\cal L}_{\rm eff} = {\cal L}_{\rm SM} + \sum_{n,\ D \geq 5} \frac{C^{(D)}_n}{ \Lambda_{NP}^{D-4} } \, O^{(D)}_n ~.
 \label{eq:SMEFT}
 \end{equation}
The Wilson Coefficients $C_n^{(D)}$ encode additional model information (couplings, ratios of masses, etc). If the underlying model is known, the Wilson coefficients can be calculated in terms of the model parameters, so the  effective Lagrangian describes the low-energy limit of any weakly-coupled extension of the SM containing only heavy New Particles. The leading CLFV operators appear at dimension $D=6$ and are therefore suppressed by $1/\Lambda_{NP}^2$. 

The EFT framework  is applicable to processes in which the center-of-mass energy is well below the expected scale of NP. This means that lepton and meson decays can be analyzed in this framework, as can intermediate-energy colliders such as the EIC  with center-of-mass energy $\sqrt{s} < v \sim 200$~GeV. Moreover, given the null results so far for NP searches at the LHC, the SMEFT also can be applied, with some caveats, to the analysis of LHC processes, as performed in~\cite{HeavyStates}.

\begin{figure}[htb]
 \centering
 \includegraphics[width=0.7\columnwidth]{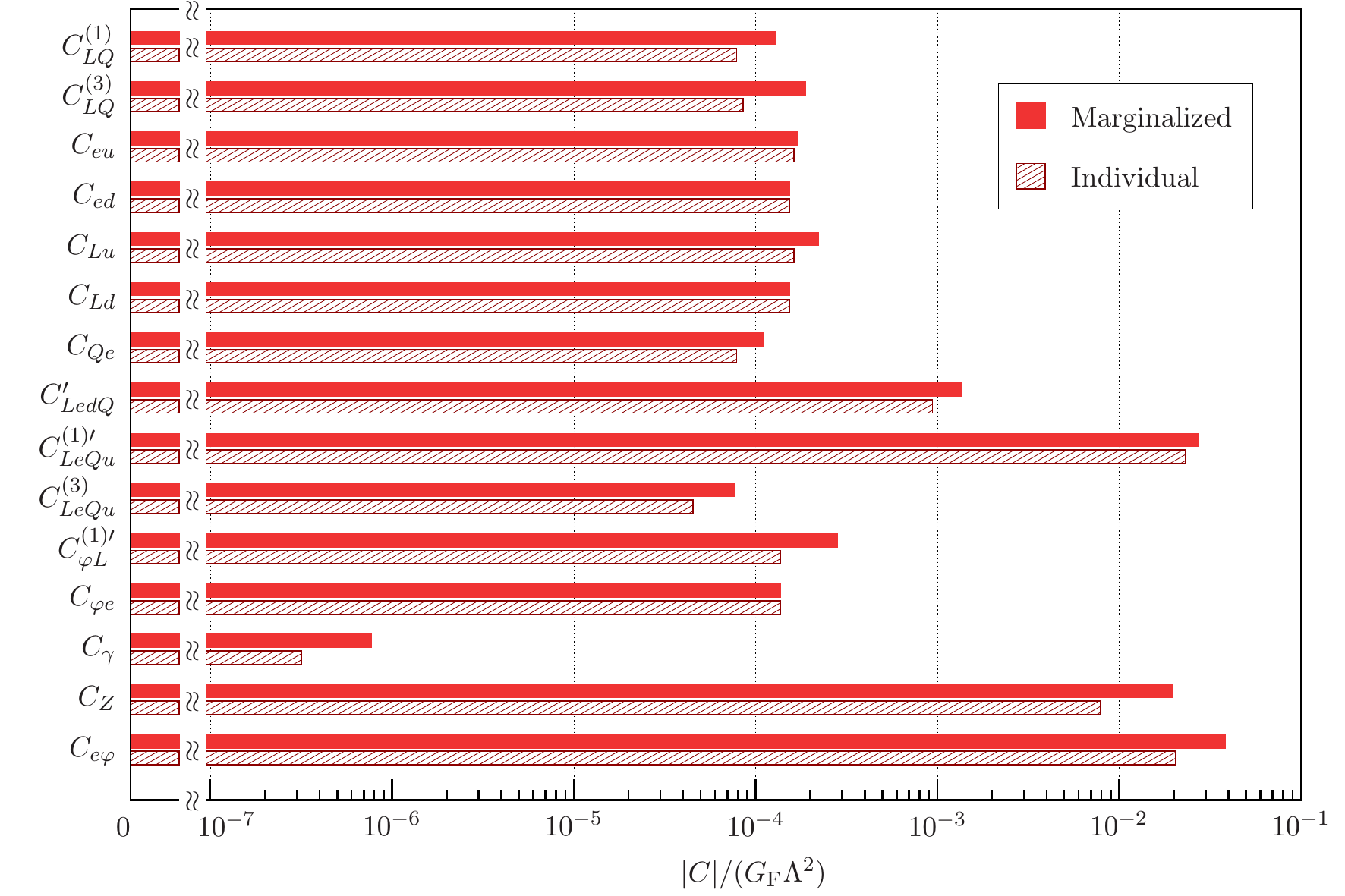}
 \caption{Allowed values for $C/(G_\text{F}\Lambda^2)$ based on the expected Belle-II limits, comparing the individual and marginalized analyses for hadronic tau decays, given at the 99\% confidence level.}
 \label{fig:6_Belle_2}
\end{figure}

CLFV processes involving $\tau$ leptons were studied in an EFT framework in~\cite{Banerjee:2022xuw} (see also ~\cite{Davidson:2020hkf,Cirigliano:2021img,Antusch:2020vul,Husek:2020fru,Gninenko:2018num,Takeuchi:2017btl,Hazard:2016fnc,Celis:2014asa,Celis:2013xja,Petrov:2013vka,Daub:2012mu,Han:2010sa,Gonderinger:2010yn,Dassinger:2007ru,Matsuzaki:2007hh,Black:2002wh}); in addition, $(\mu\to \tau) \times (\tau \to e)$ interactions can contribute to $\mu\to e$ processes~\cite{Ardu:2022pzk}. The constraints on Wilson coefficients resulting from the non-observation of CLFV at Belle-II are illustrated in Figure \ref{fig:6_Belle_2}, taken from~\cite{Banerjee:2022xuw}; the ``marginalized'' limits are constraints in the presence of all the coefficients, and are comparable to the ``individual'' bounds that apply to one operator at a time. This illustrates the ability of $\tau$ decays to distinguish among coefficients and thereby among models. 

\subsection{Models}

The above EFT approach is model agnostic and can capture the effects of any \emph{heavy} new physics model. However, the large number of baryon-number-conserving CLFV SMEFT operators -- 888 at mass dimension six~\cite{Fonseca:2019yya} -- can cause dismay.
The study of simple well-motivated models provides a complementary approach: if the number of new fields and parameters is small, 
CLFV rates can be predicted. Neutrino-mass models are of particular interest since they necessarily violate lepton flavor.


In the celebrated type-I seesaw mechanism~\cite{Minkowski:1977sc,Yanagida:1979as,Gell-Mann:1979vob,Mohapatra:1979ia}, heavy right-handed neutrinos with diagonal mass matrix $M_R$ generate a Majorana neutrino mass matrix $M_\nu \simeq - \tfrac12 v^2\,y_D M_R^{-1} y_D^T$ as well as $d=6$ CLFV operators $(y_D M_R^{-2} y_D^\dagger)_{\alpha\beta}\overline{L}_\alpha H i\slashed{\partial} H^\dagger L_\beta$~\cite{Broncano:2002rw} that induce $\ell_\alpha\to \ell_\beta \gamma$ and more. Despite the close connection, even full knowledge of $M_\nu$ does not fix the relevant CLFV matrix $y_D M_R^{-2} y_D^\dagger$~\cite{Casas:2001sr,Davidson:2002qv,Broncano:2003fq}, i.e.~there are no definite CLFV predictions; on the flip side, CLFV processes can be sizable here~\cite{Antusch:2006vwa,Abada:2007ux,Coy:2018bxr} and any observation would provide crucial complementary information about the seesaw mechanism~\cite{Broncano:2003fq}. The desired flavor structure that enhances CLFV while keeping $M_\nu$ small can naturally be obtained in extensions such as the inverse seesaw mechanism~\cite{Wyler:1982dd, Mohapatra:1986aw, Mohapatra:1986bd, Gonzalez-Garcia:1988okv, Abada:2014vea}.

Other neutrino-mass models have more predictive power. For example, in the type-II seesaw mechanism~\cite{Konetschny:1977bn, Magg:1980ut, Schechter:1980gr, Cheng:1980qt, Mohapatra:1980yp}, an $SU(2)_L$-triplet scalar obtains a vacuum expectation value that generates a Majorana neutrino mass matrix $M_\nu$. The charged triplet components induce CLFV processes with flavor structure directly connected to $M_\nu$ and the oscillation angles and neutrino masses residing inside~\cite{Pich:1984uoh,Ma:2000xh,Kakizaki:2003jk,Chakrabortty:2012vp}. While the overall CLFV rates depend on the unknown triplet masses, \emph{ratios} of CLFV channels only depend on neutrino parameters and can be predicted.
For example, the rate for $\ell_\alpha\to \ell_\beta\gamma$ is proportional to $|(M_\nu M_\nu^\dagger)_{\beta\alpha}|^2$, which is completely specified by neutrino-oscillation data and predicts~\cite{Chakrabortty:2012vp}
\begin{align}
\text{BR} (\tau\to \mu \gamma) \simeq 23\, \text{BR} (\tau\to e \gamma)\simeq 3.5\, \text{BR}(\mu\to e\gamma)\,.
\end{align}
The rates for $\ell_\alpha\to\ell_\beta\bar{\ell}_\eta\ell_\sigma$ depend on the individual $M_\nu$ entries and hence Majorana CP phases and the absolute neutrino mass scale, neither of which are known yet~\cite{Dinh:2012bp}; in this model, observations of several CLFV modes could therefore provide information about these difficult-to-measure neutrino parameters, in addition to the triplet masses. CLFV in the $\tau$ sector comes with larger branching ratios than in the $\mu$ sector, which is however more than compensated by the superior experimental reach for $\mu$ CLFV.

Models that generate neutrino masses at loop rather than tree level naturally require lower new-physics scales and thus enhanced CLFV~\cite{Cai:2017jrq}. However, even with the flavor structure fixed or linked to charged-fermion mixing, the overall CLFV rates cannot be predicted without knowledge of the new physics scale or masses. 
This could be achieved in models that explain current anomalies such as $(g-2)_\mu$~\cite{Bennett:2006fi, Muong-2:2021ojo}, lepton-flavor non-universality in $B$ decays~\cite{LHCb:2017smo,LHCb:2021trn}, or CDF's $M_W$ mass measurement~\cite{CDF:2022hxs}, all of which can be resolved with new physics around the electroweak scale that generically yields large CLFV. The CLFV constraints are often so stringent that it appears more natural to suppress or even eliminate CLFV altogether via symmetries/flavor-alignment~\cite{Hambye:2017qix, Davighi:2020qqa, Greljo:2021xmg, Greljo:2021npi, Faroughy:2020ina, Isidori:2021gqe}, Minimal Flavor Violation~\cite{Chivukula:1987py, DAmbrosio:2002vsn, Cirigliano:2005ck, Davidson:2006bd} being a well-known example. Should any of the current anomalies survive and prove the existence of TeV-scale physics beyond the SM, it is generically expected to lead to CLFV as well. More detailed discussions of CLFV models and their connection to other issues can be found in ~\cite{Kuno:1999jp, Raidal:2008jk, deGouvea:2013zba, Lindner:2016bgg, Calibbi:2017uvl}.

Models involving \emph{heavy} particles effectively reduce the number of SMEFT operators to something manageable or even predictive. Models involving \emph{light} particles cannot be described by the SMEFT at all and thus require dedicated analyses~\cite{LoI080} or extensions of the SMEFT by additional light particles~\cite{Georgi:1986df, Brivio:2017ije}. For example, the majoron $J$~\cite{Chikashige:1980ui,Schechter:1981cv} as the  Goldstone boson of lepton number often appears in neutrino-mass models and leads to the CLFV processes $\ell_\alpha\to\ell_\beta J$~\cite{Pilaftsis:1993af, Garcia-Cely:2017oco}. Pseudoscalars~\cite{Kim:1986ax, Calibbi:2020jvd} such as axions and familons can similarly induce such CLFV, as can $Z'$ gauge bosons~\cite{Foot:1994vd, Heeck:2016xkh}. Depending on the lifetime and decay channels of the light particle, even displaced-vertex signatures such as $\ell_\alpha\to\ell_\beta J\to \ell_\beta e^+ e^-$~\cite{Heeck:2017xmg,Cheung:2021mol} are possible and require dedicated searches.

\section{Muon experimental overview}

The muon has consistently provided powerful constraints on CLFV reactions (see e.g.~\cite{Calibbi:2017uvl, Bernstein:2013hba}). Muons have a relatively long lifetime and a limited number of decay channels, resulting in much simpler final states than those of the heavier tau lepton. In addition, intense muon beams are available at several facilities, providing the high statistics needed to study processes with extremely small rates. Three main transitions have been investigated so far: $\meg$, $\meee$ and $\mec$ conversion in the Coulomb field of a nucleus, in addition to muonium-antimuonium oscillations. Rare muon decay experiments typically exploits the kinematical constraints of a decay at rest from intense positive muon beams stopped in a target. These studies are only possible with positive muons, since negative muons would be captured by the target nuclei, distorting the decay kinematics. The most recent experiments exploit surface muon beams ($p_\mu = 29.8 \MeV$) produced by the decay of pions at rest on the surface of a production target. By contrast, muon conversion experiments stop negative muons in matter, wherein muons are captured in atomic orbits before undergoing the conversion process. The reach and complementarity of the three transitions is illustrated in Fig.~\ref{fig:NPreach}. Current measurements probe NP mass scales at the level of $10^3 - 10^4 \TeV$ over a large fraction of parameter space, and future experiments will increase the sensitivity by an order of magnitude. 


This section first reviews the current landscape of muon CLFV experiments and their expected performance. A staged program of future next-generation experiments and facilities is then discussed. Mu2e-II is proposed as a near-term, low cost follow-on to Mu2e, extending the investigation of muon to electron physics by an order of magnitude or more in sensitivity. By leveraging existing investments in Mu2e and PIP-II, Mu2e-II plans to starts its construction phase before the end of the decade. On a longer term, the Advanced Muon Facility is a more ambitious proposal for a new high-intensity muon complex at FNAL. This facility would provide the world's most intense positive and negative muon beams, enabling a whole suite of experiments and synergies with a light dark matter search program.

\subsection{Muon Flavor Violation Experiments in this Decade}

\subsubsection{Muon decay experiments}
The search for $\meg$ is based on the reconstruction of a positron and a photon emitted back-to-back from the stopping target, each with an energy of $ 52.8~\MeV$. The background rejection can independently exploit information on the positron and photon. At very high muon stopping rates, accidental backgrounds from pile-up largely dominate over the intrinsic $\menng$ background. Since accidental backgrounds are proportional to the square of the beam intensity, the experimental sensitivity plateaus once a threshold muon stopping rate is crossed.

The best limit on $\meg$ has been set by the MEG experiment at PSI, $\rm BR(\mu^+ \to e^+ \gamma) < 4.2 \times 10^{-13}$ at 90\% confidence level~\cite{MEG:2016leq}. The positron kinematics are reconstructed with a drift chamber in a graded solenoidal magnetic field. The photon energy, time and production point are determined with a LXe detector instrumented with PMTs. The MEG experiment has recently been upgraded with a new drift chamber, silicon photomultipliers at the entrance of the LXE detector, and a more granular positron timing detector~\cite{MEGII2018}. The upgraded experiment aims for a final sensitivity of $\sim 6 \times 10^{-14}$ after three years of data taking~\cite{Meucci:2022qbh}. On-going studies show that that incremental improvements in photon calorimetry and positron tracking could push the limit below $10^{-14}$~\cite{next_meg}, but fully exploiting beam rates of $10^{10}~\mu$/s or more and breaking the $10^{-15}$ barrier will require a conceptually new experimental approach.

The $\meee$ decay is reconstructed by combining three tracks originating from the same position. The dominant background arise from $\mu^+ \to e^+ e^+ e^- \overline \nu_\mu \nu_e$ decays since the accidental background (coincidence of one or more standard Michel decays with a positron produced from Bhabha scattering or radiative decay) is strongly suppressed up to very high muon beam intensities. All backgrounds can be controlled using vertexing and kinematic requirements, and the accidental contribution can be further reduced by coincident timing requirements.

The current limit has been set by the SINDRUM experiment at PSI, $\rm BR(\mu^+ \to e^+ e^+ e^-) < 1.0 \times 10^{-12}$ at 90\% CL~\cite{SINDRUM:1987nra}.  The Mu3e experiment at PSI~\cite{Mu3e:2020gyw} plans to improve the sensitivity by several orders of magnitude. The experimental apparatus contains four layers of High-Voltage Monolithic Active Pixel Sensors (HV-MAPS) surrounding a muon stopping target, scintillating fibres and scintillating tiles for a precise timing of the charged tracks. A Phase-I experiment is planned on the same beam line as the MEG experiment to reach a sensitivity of $\sim 10^{-15}$. A Phase-II detector with additional tracking and timing capabilities would take full advantage of the high muon flux foreseen at the proposed HIMB facility at PSI~\cite{Aiba:2021bxe} to improve the sensitivity by an additional order of magnitude. 

In addition to searches for $\meg$ and $\meee$, decay experiments can also investigate LFV reactions of the type $\meX$ or $\megX$, where $X$ denotes a new neutral particle escaping undetected or decaying into into SM fields. Examples of such particles include LFV axion-like particles~\cite{Calibbi:2020jvd}, familons~\cite{PhysRevLett.49.1549} or dark photons~\cite{Echenard:2014lma}. The $\meX$ decay is characterized by the emission of a mono-energetic positron, which can be identified as a narrow peak on top of the $\mu^+ \to e^+ \overline \nu_\mu \nu_e$ energy spectrum. Polarized muons could potentially boost the sensitivity if the NP decays are controlled by a different structure than that of the weak interaction (e.g.\ $V+A$)~\cite{Calibbi:2020jvd}. The strongest limit on this decay have been set by Jodido {\it et al.} at TRIUMF~\cite{jodidio} and the TWIST experiment~\cite{twist_X} with bound at the level of $10^{-6} - 10^{-5}$. Dedicated experiments at very high muon beam rates could significant improve these constraints~\cite{Calibbi:2020jvd}. Visible decays could be explored in inclusive searches, such as $\mu^+ \to e^+ X$ with either $X \to \gamma \gamma$ or $X \to e^+ e^-$. The first channel was recently investigated by MEG~\cite{meg2g}, while the second could be explored at Mu3e. Similarly, $\mu^+ \rightarrow e^+ \gamma \gamma$ decays could studied with a setup similar to MEG-II, providing sensitivity to couplings poorly probed by other muon CLFV searches. 

\subsubsection{Muon conversion experiments}
The search for $\mu^-N - e^-N$ conversion is carried out by stopping a negative muon beam in a target, wherein muons are captured in atomic orbits and converted into electrons through a coherent interaction with the nucleus. Since the nucleus remains unchanged during this process, the energy of the outgoing electron is close to the muon rest mass. The rate of the conversion process relative to ordinary muon capture is conventionally defined as:
$$R_{\mu e} = \frac{\Gamma(\mu^- + N(A,Z) \rightarrow e^- + N(A,Z))} {\Gamma(\mu^- + N(A,Z) \rightarrow \rm all \, captures)}$$ where $N(A,Z)$ denotes the mass and atomic numbers of the target nuclei. Incidentally, measurements of negative muon capture rates provide valuable inputs in the calculation of the amplitudes of the virtual transitions in neutrinoless double beta decays~\cite{Suhonen:2006bh}.

The current experimental bound has been set by the SINDRUM-II experiment at PSI using a Au target, $R_{\mu e} < 7\times10^{-13}$ at 90\% CL~\cite{SINDRUMII:2006dvw}. The Mu2e experiment~\cite{bartoszek2015mu2e}, under construction at Fermilab, aims to reach a single event sensitivity on the conversion rate on an Al target of $\sim 3\times10^{-17}$. The COMET collaboration at J-PARC will proceed in two phases, with a planned sensitivity of ${\cal O}(10^{-15})$ on Al for Phase-I~\cite{comet-tdr} and a similar sensitivity to Mu2e for Phase-II~\cite{COMET:2009qeh,Krikler:2016qij}. 
Both Mu2e and COMET use pulsed proton beams to form an intense muon beam, transported onto a stopping target. The conversion electron is reconstructed with a high-resolution tracking system and a calorimeter placed in a solenoid. The muon transport line is based on curved solenoids to shield the detector from the direct line of sight of the production target and select negatively charged muons. In the Mu2e configuration, the stopping target is located directly in front of the detector. This design has the advantage of being charge symmetric, enabling the search for $\mu^-N  \rightarrow e^+ N'$ decays and measure positrons from radiative pion captures, but the innermost regions of the detector must be left uninstrumented to withstand the large flux of low-momentum particles. By contrast, COMET uses an additional curved solenoid downstream of the stopping target before a tracking detector to limit the acceptance to electrons with momenta near that of the expected signal. This allows for a fully instrumented volume, but at the cost of charge symmetry. 

The DeeMe experiment at J-PARC~\cite{Teshima2019} has adopted a different scheme, using a single target to produce and capture muons. Pulsed proton bunches are transported to a graphite target, and a fraction of the muons produced by pion decays will be captured near the surface of the target itself and form muonic atoms. Electrons from the conversion process are transported by a secondary beamline to a compact magnetic spectrometer. The experiment is currently in preparation and has a projected sensitivity of ${\cal O}(10^{-13})$.

\subsubsection{Muonium-antimuonium oscillations}
Muonium ($M$) is a bound state of a positive muon and an electron, with a lifetime similar to that of the muon. The spontaneous conversion of muonium into antimuonium via $\Delta L = 2$ interaction opens the possibility to study muonium – antimuonium oscillations. Theoretical analyses of the conversion probability have been performed both in particular NP models (see e.g.~\cite{Clark:2003tv,Li:2019xvv}) and in effective field theory~\cite{Conlin:2020veq}. 

Muonium atoms are usually formed by injecting and slowing down a surface $\mu^+$ beam in material. A fraction of the positive muons could spontaneously capture an electron and emerge in vacuum as muonium. After conversion, the antimuonium is identified by reconstructing the Michel electron from the muon decays and the shell positron. Based on this approach, the MACS experiment at PSI set a bound on the $M - \bar{M}$ conversion probability ($P_{M \bar{M}}$) of $P_{M \bar{M}} < 8.3 \times 10^{-11}$ at 90\% CL~\cite{Willmann:1998gd}.

The MACE experiment~\cite{Bai:2022sxq} has been proposed to improve this sensitivity by more than two orders of magnitude. The proton beam required to produce the surface muon beam could be provided by the China Spallation Neutron Source or the continuous proton beam of the China Initiative Accelerator Driven sub-critical System. The rate of muonium formation and diffusion in vacuum could be enhanced by replacing the silica powder used by the MACS experiment with a laser-ablated silica aerogel target. The signal would be identified by reconstructing the Michel electron with a magnetic spectrometer and the shell positron with a composite detector system  comprising a microchannel plate (MCP) and an electromagnetic calorimeter. The shell positron is accelerated to an energy around a few $\keV$ and guided to the detector system through a transport line before annihilating into a photon pair in the MCP. The coincidence between the Michel electron, the shell positron signal in the MCP and the two photons in the calorimeter is used to suppress the various backgrounds. 

An alternative scheme based on a surface muon beam crossing a layer of superfluid helium has also been proposed~\cite{Adelberger:2022sve}. This beam could be produced by either the existing 400-MeV Linac at FNAL, or the PIP-II accelerator under construction. Slow antimuons would then be directed into a small cryostat cooled to sub-Kelvin temperature, and form muonium in a layer of superfluid helium. The muonium atoms are ejected vertically from the upper superfluid helium surface, producing a quasi-monoenergetic, quasi-parallel muonium beam in vacuum. This beam would enable (especially at PIP-II) world-leading sensitivities for muonium gravity, muonium spectroscopy, and muonium – antimuonium oscillation experiments (see the "Fundamental Physics in Small Experiments" Topical Group report~\cite{rapportTG3} for more details).

\subsection{Future initiatives and next-generation facilities}

\subsubsection{Mu2e-II}
Mu2e-II~\cite{Mu2e-II:2022blh} is a proposed evolution of the Mu2e experiment at FNAL with the aim of increasing the sensitivity of $\mec$ conversion by an order of magnitude or more over Mu2e. By reusing as much of the current infrastructure as possible, construction of this upgrade is planned to start before the end of the decade. Should a conversion signal be observed, Mu2e-II could investigate the underlying physics by measuring the conversion rate for mid-Z target materials, such as Ti or V~\cite{Cirigliano:2009bz}. Alternatively, increasing the sensitivity to probe higher NP mass scales will be required if no signal is seen at the currently planned experiments. Mu2e-II would also investigate the $\Delta L=2$ process $\mu^- N \rightarrow e^+ N$ with increased sensitivity, and the precise determination of the tail of the muon decay-in-orbit spectrum could further explore NP signatures~\cite{GarciaTormo:2011jit,Uesaka:2020okd}. 

Mu2e-II will exploit the PIP-II linac at Fermilab~\cite{Ball2017}, currently under construction, to provide the intense 800 MeV proton beam (100 kW on target) needed for the experiment. The gain in sensitivity over Mu2e is achieved by a combination of higher intensity and duty factor, since the muon production rate are comparable at 800 MeV and 8 GeV for a given beam power. Handling a more powerful beam while keeping the background under control presents several challenges. The PIP-II linac pulses are narrower than the Mu2e resonantly extracted beam, improving the beam extinction, but a section of the beamline, the production solenoid, and associated shielding may need significant changes to handle the higher beam power. On the other hand, the proton beam energy is below the antiproton production threshold, eliminating one potential source of background.

While the detector layout remains essentially unchanged compared to Mu2e, the higher occupancy and background rejection requirements place stringent constraints on the sub-system performance. In particular, improvements in tracker resolution and in pattern recognition are required to adequately reject the muon decay-in-orbit background. The design of the current Mu2e tracker, a straw tube chamber with 15 $\mu$m aluminized mylar straws, would need to be revisited. A R\&D program is underway to explore decreasing the thickness of the straws to reduce multiple scattering~\cite{Ambrose2020a}. The Mu2de calorimeter, used for triggering, particle identification and validating the tracker momentum measurement, is composed of CsI crystals read out by SiPMs. The CsI scintillation light decay time ($\sim 30-40$ ns) is marginal for Mu2e-II, and alternatives are being investigated. BaF$_2$ is an especially promising candidate, with a very fast (sub-nanosecond) component around 220 nm. A R\&D effort has started to suppress the slow component at longer wavelength by doping with yttrium and developing a UV-sensitive, solar-blind readout~\cite{Davydov2020,Hitlin2020a, Hu2020}. Cosmic rays are also a major background and their rejection is crucial to achieve the desired sensitivity. The higher running duty factor will increase the live time by about a factor of three compared to Mu2e, and the scintillator-based Mu2e cosmic ray veto (CRV) system needs to be upgraded. A new geometry of the CRV counters to reduce the detection inefficiency is under study, as well as other technologies (such as RPCs)~\cite{Byrum2020}. Mu2e-II will also have an order of magnitude higher data rate than Mu2e, posing challenges for the trigger and data acquisition system, and potential approaches to mitigate these issues are also under investigation~\cite{Gioiosa2020}.

\subsubsection{Advanced Muon Facility at FNAL}
A more ambitious proposal, the Advanced Muon Facility (AMF), would exploit the full potential of the PIP-II accelerator to explore muon physics with unprecedented sensitivity~\cite{CGroup:2022tli}. A suite of experiments could be pursued at this complex, from charged lepton flavor violation and muonium-antimuonium oscillations to muon EDM and muon spin rotation. In addition, this program has many synergistic activities with R\&D efforts to develop a muon collider, as well as a future beam dump experiment searching for dark matter and other light NP at FNAL~\cite{boosterDM,toupsDM} (see the "Dark Sector Studies at High Intensities" Topical Group report~\cite{rapportTG6}  for more details). Dedicated CLFV experiments at AMF could potentially improve the sensitivity of decay channels by two orders of magnitude compared to the ultimate rates probed by currently planned initiatives, and reach conversion rates down to the level of $10^{-18}$ with a proton beam power of ${\cal O}(100)$ kW, and $10^{-19}$ or lower with ${\cal O}(1)$ MW. This complex would also enable the study of muon conversion with high-$Z$ target materials, which could provide critical information about the nature of the underlying NP~\cite{Cirigliano:2009bz}. 

\begin{figure}[htb]
\begin{center}
\includegraphics[width=0.9\textwidth]{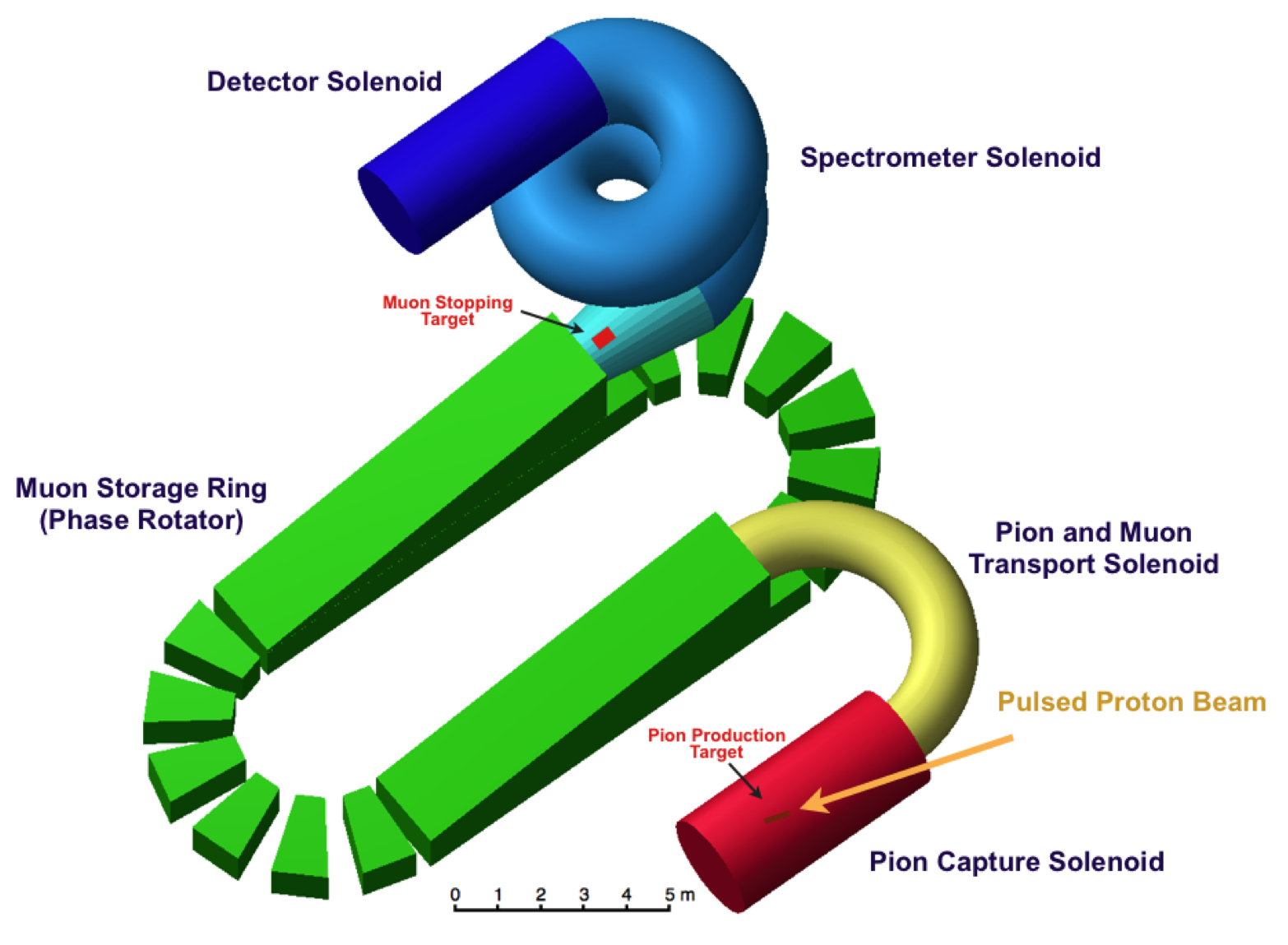}
\end{center}
\caption{The PRISM concept, adapted from~\cite{KUNO2005376}, showing the facility configured for muon conversion experiments. Not shown are the PIP-II linac, the RF beam splitter and transport lines, the compressor ring, and the induction linac. The spectrometer and detector solenoids could be replaced for upgrades or new, different experiments.}
\label{fig:prism}
\end{figure}

The Advanced Muon Facility is based on a small fixed-field alternating gradient synchrotron (FFA), used to produce an intense muon beam with well-defined momentum from the PIP-II accelerator. The PRISM (Phase Rotated Intense Source of Muons) system~\cite{KUNO2005376}, shown in Fig.~\ref{fig:prism}, provides a reference concept. Short high intensity proton bunches are delivered to a production target surrounded by a capture solenoid with a field at about 5T, well within current capabilities. The muons produced by pion decays are then injected into the FFA ring by a transport system. The phase rotation decreases the momentum spread of the muons, trading momentum spread for time spread. During the RF phase rotation, the remaining pion contamination is reduced to negligible levels. A cold quasi-monochromatic muon beam is then extracted to the detector system. The feasibility of the FFA approach was demonstrated with a dedicated prototype at the Research Center of Nuclear Physics (RCNP) of Osaka University~\cite{Witte:2012zza}. 

The realization of this concept presents several challenges that must be addressed through a dedicated R\&D program. The design of the proton compressor ring for a 1~MW facility is very complicated to achieve from the $0.8 \GeV$ PIP-II beam with conventional magnets. However, a 50-200~kW AMF facility would be consistent with a more conservative set of parameters described in~\cite{toupsDM} for a $0.8-1.2\GeV$ ring termed ``C-PAR’’. The proton compressor for a 1~MW AMF facility would likely require a higher injection energy ($2-3\GeV$) or alternatively use of super-ferric magnets. Another major limiting factor is the beam power that can be absorbed by the target. Several concepts have been developed to handle a beam power of ${\cal O}(100)$ kW in a capture solenoid, but additional efforts are needed to design a MW-class target. A similar challenge is faced by the muon collider~\cite{calvianiTalk}; the two projects are potentially synergistic, with the target required for this program representing both a staging and an R\&D platform for the demands of the muon collider. Finally, the beam dynamics in the muon transport solenoid and the FFA ring are significantly different. The design of an efficient beam transport and injection system will need detailed studies to be validated and optimized. 

Improving the $\mec$ sensitivity could be accomplished by extending the COMET approach to a larger spectrometer to reduce backgrounds in the detector, as proposed by the PRISM/PRIME experiment~\cite{KUNO2005376}, together with a high-resolution, low-mass tracking system. A number of promising detector technologies have been identified for the latter, including the proposed Mu2e-II straw-tracker with 8 $\mu$m wall thickness~\cite{Mu2e-II:2022blh}, or low-mass silicon sensors such as HVMaps or micro-pattern gas detectors proposed for the Belle-II tracking TPC~\cite{Forti:2022mti}.  

AMF could be implemented in a phased approach: a conceptual design for a complex with a $\sim 100$ kW proton beam power using PIP-II as a first stage, followed by an upgrade to reach a final $\sim 1$ MW power. A 100 kW facility would already provide significant sensitivity improvements for both decay and conversion experiments, and allow measurement of conversion in high-$Z$ materials, currently inaccessible with pulsed-beam experiments. Starting the design studies in the near future would allow the realization of this program shortly after the completion of the Mu2e experiment on the FNAL site, and operate simultaneously with LBNF/DUNE. 

The FFA cannot provide the ideal beam structure for muon decay experiments -- a continuous positive muon beam -- as the large size of the FFA beam makes slow extraction impractical. Instead, we envision a standard surface muon beam created by coupling the output of PIP-II to the same production solenoid and target, selecting positive muons instead of the negative muons for the capture experiments. Based on the operation of the $\pi$E5 beamline at PSI, a proton beam power of $\sim 100 \rm \, kW$ would yield a $10^{12}~\mu^+$/s beam, roughly two orders of magnitude larger than the proposed HiMB facility at PSI~\cite{Aiba:2021bxe}. We are also investigating whether this surface muon beam could be further slowed in an induction linac; a lower energy beam stops in a shorter distance in material, improving the vertex resolution required for both decay experiments, which will be limited by accidental backgrounds. This facility could also be used to study muonium physics~\cite{Kaplan:LOI}. As discussed in the previous section, improvements in photon calorimetry and track reconstruction could push the sensitivity of $\meg$ searches close to $10^{-15}$ with the HiMB at PSI~\cite{next_meg}. Similarly, the current approach for $\meee$ plateaus near $10^{-16}$ with the current technology~\cite{Hesketh:2022wgw}. Conceptually new experimental approaches are required to take full advantage of a bright surface muon beam powered by PIP-II.
\section{Tau experimental overview} 
In contrast to muon CLFV searches, in which a given decay is usually studied by a dedicated experiment, tau CLFV searches can be conducted over many final states with large data sets collected at $e^+e^-$ or hadron colliders. In addition, the large $\tau$ mass greatly decreases the GIM suppression, enhancing the signal rate with respect to the corresponding muon channel in some scenarios. However, the typical sample size collected at colliders is much smaller than the muon production rate at dedicated facilities, partially negating this advantage. The experimental landscape will undergo tremendous progress in the next decade, with Belle-II and LHC working towards collecting large data sets. The Electron-Ion Collider (EIC), the Super $\tau$-Charm Facility (STCF), and the Future Circular Collider (FCC) could also play a major role on a more distant horizon. This section reviews the efforts in turn.

\subsection{Tau Flavor Violation Experiments in this Decade}
The first generation of $B$-factories collected large data samples of $\tau$ pairs produced in $e^+e^- \rightarrow \tau^+\tau^-$ events. The current bounds probe branching fractions at the level of $10^{-8} - 10^{-7}$ for many purely leptonic decays, as well as final states with one or several hadrons~\cite{HFLAV:2019otj}. 

The Belle-II experiment at SuperKEKB~\cite{Belle-II:2010dht} is expected to collect a data set $\sim30\times$ larger than the combined BABAR and Belle integrated luminosities. Together with increased reconstruction efficiencies, this experiment could improve the current limits by two orders of magnitude for the cleanest channels (e.g $\tau \rightarrow 3\ell$), and an order of magnitude for mode with irreducible backgrounds~\cite{Belle-II:2018jsg}, as shown in Fig.~\ref{fig:tauLFV}. Additional gain would be possible by upgrading SuperKEKB to provide polarized electron beams with approximately 70\% polarization~\cite{BellePolarization}. In that mode of operation, the helicity angles of the $\tau$ pair decay products can be used to further suppress the background, resulting in a sensitivity increase of the $\tau \rightarrow \mu \gamma$ channel of $\sim10\%$~\cite{BellePolarization}. A similar gain is expected for the $\tau \rightarrow e \gamma$ reaction. More interestingly, the Dalitz plot of the polarized $\tau^- \rightarrow \mu^- \mu^+\mu^-$ decay can be used to infer the Lorentz structure of the CLFV coupling, should a signal be observed~\cite{Matsuzaki:2007hh,Dassinger:2007ru}. 
Tau decays also allow for precision tests of charged lepton flavor universality (LFU), the assumption that lepton coupling to charge gauge bosons of the electroweak interaction have equal strength. These tests typically consist of precise measurements of branching-fractions ratios, such as ${\cal B}(\tau \rightarrow \mu \bar{\nu}_\mu \nu_\tau) / {\cal B}(\tau \rightarrow e \bar{\nu}_e \nu_\tau)$ or ${\cal B}(\tau \rightarrow K^- \nu_\tau) / {\cal B}(K \rightarrow \mu^- \nu_\mu)$ (see e.g.~\cite{LFU}). Belle-II will significantly improve the precision on many inputs to measure LFU quantities, yielding some of the most stringent constraints on non-SM deviations from charged current lepton universality~\cite{Belle2WP}.

At the LHC, $\tau$ leptons are produced almost entirely from $b$ and $c$ hadron decays. The CMS and LHCb collaborations have taken advantage of the large inclusive $\tau$ production cross-section to search for $\tau^- \rightarrow \mu^-\mu^+\mu^-$ decays, setting limits in the range $4.6-8 \times 10^{-8}$~\cite{LHCb:2014kws, CMS:2020kwy, ATLAS:2016jts}. Bounds in the vicinity of $10^{-5}$ have also been derived for lepton flavor violating $b$-hadrons decays in $B^0 \rightarrow \mu^\pm \tau^\mp$, $B^0_S \rightarrow \mu^\pm \tau^\mp$ and $B^+ \rightarrow K^+ \mu^- \tau^+$ by LHCb~\cite{Aaij:2019okb,LHCb:2020khb}. In the HL-LHC era, the LHCb Upgrade II detector plans to collect $300 \ifb$ of data at $14 \TeV$, opening the possibility to probe the $\tau^- \rightarrow \mu^- \mu^+\mu^-$ branching fraction at the level of $10^{-9}$~\cite{LHCb:2018roe}. A similar level of sensitivity is expected at the ATLAS and CMS experiments with a data set of $3000 \ifb$~\cite{atlast3m,cmsupt3m}.

\begin{figure}[htb]
\begin{center}
\includegraphics[width=0.8\textwidth]{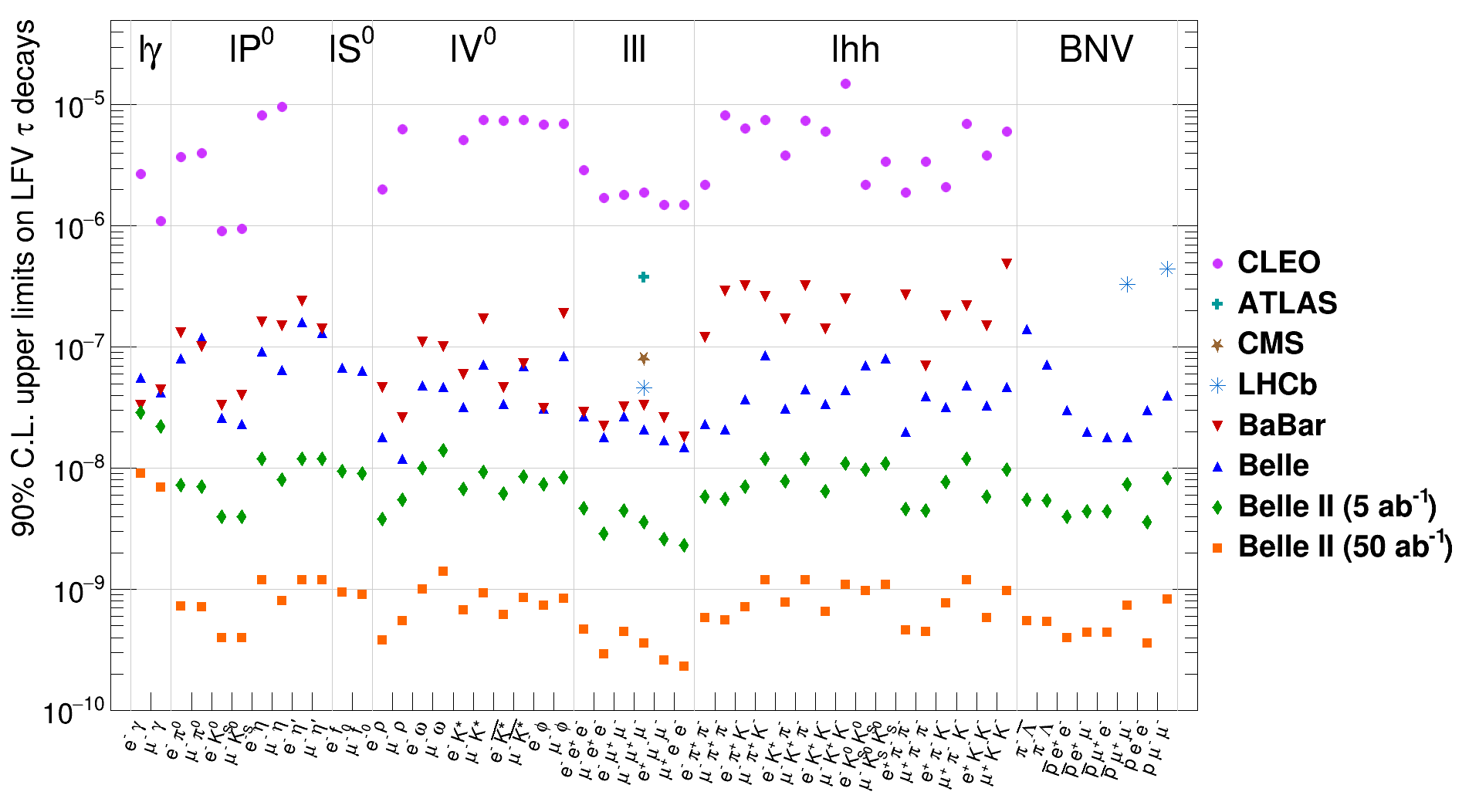}
\end{center}
\caption{Projection (by simple statistical scaling) of expected upper limits at the Belle-II experiment~\cite{Belle2WP} and current status of observed  upper limits at CLEO, BABAR, Belle, ATLAS, CMS and LHCb experiments~\cite{Amhis:2019ckw} on LFV, LNV and BNV processes in $\tau$ decays.}
\label{fig:tauLFV}
\end{figure}

\subsection{Future initiatives and next-generation facilities}

The Super $\tau$-Charm Facility~(STCF)~\cite{STCF} is a proposed symmetric electron-positron collider designed to operate at $\sqrt{s}=2 \sim 7 \GeV$ with a peak luminosity of $0.5\times10^{35} \rm \, cm^{-2}s^{-1}$ or higher, with a possible upgrade including a polarized electron beam~\cite{Luo:2019gri}. This facility could produce $3.5 \times 10^{9}$ $\tau^-\tau^+$ pairs at $\sqrt{s}=4.26 \GeV$ per year, and $\sim 10^8$ $\tau$-pairs near the production threshold. While the production rate is lower in the latter case, operating near threshold could offer better control of systematic uncertainties by collecting data just below this energy. 
The sensitivity of two benchmark CLFV processes, $\tau^- \rightarrow \mu^-\mu^+\mu^-$ and $\tau^-\rightarrow \mu^-\gamma$, has been studied at $\sqrt{s}=4.26 \GeV$~\cite{STCF:clfv}. The signal region is almost background-free after applying the signal selection procedure for the $\tau^- \rightarrow \mu^-\mu^+\mu^-$ channel, yielding a sensitivity of $1.4\times10^{-9}$ at 90\% CL for $3.5\times10^{9}$ $\tau$-pairs. The situation is slightly more challenging for the $\tau^-\rightarrow \mu^-\gamma$ final state due to the presence of a larger background arising from photon and $\pi/\mu$ mis-identification. Several approaches have been explored to suppress these background, yielding sensitivity estimates in the range $(1.2\sim 1.8)\times10^{-8}$ at 90\% CL.

The Electron Ion Collider~\cite{Accardi:2012qut} (EIC) is a collider designed to provide collisions between a polarized electron beam and a wide range of ions, ranging from polarized protons to unpolarized heavier ions up to uranium. The large instantaneous luminosity opens the door for precision tests of the SM. A leading observable in this arena is the electron-to-tau transition, which could be enhanced to observable levels in several BSM scenarios~\cite{Gonderinger:2010yn, Cirigliano:2021img}. The current limits on $e +  p\to \tau +X$, $\tau\to e\gamma$, and $p+p\to e+\tau + X$ have been set at HERA~\cite{ZEUS2019}, BABAR~\cite{BaBar:2009hkt}, and the LHC~\cite{ATLAS:2018mrn} respectively. A study based on the ECCE detector configuration was conducted to evaluate the reach on the $e^- \leftrightarrow \tau^-$ transition mediated by a $1.9 \GeV$ leptoquark~\cite{Banerjee:2022xuw}. Assuming 100~$\ifb$ of luminosity for the $18\times 275 \GeV$ energy configuration, the EIC should be able to improve on the limits set by HERA by up to an order of magnitude.

The FCC-ee program~\cite{Abada:2019lih,FCC:2018evy} plans to produce about $5\times 10^{12}$ $Z$ decays, out of which $1.7 \times 10^{11}$ will decay to tau pairs. This large sample will open the door to a very rich $\tau$-physics program~\cite{Pich:2020qna}, including the ability to probe the same set of LFV $\tau$ decays measured by the $B$-factories, with sensitivities in the $10^{-10} - 10^{-9}$ range. More quantitatively, a study of the $\tau^- \to \mu^-\mu^+\mu^-$ and $\tau^-\to\mu^-\gamma$ modes has been carried out~\cite{Dam:2018rfz}. The analysis relies on the identification of a tag side to select $Z \rightarrow \tau^+\tau^-$ events, searching for LFV decays in the signal side. No backgrounds were found for the $\tau^- \to \mu^-\mu^+\mu^-$ mode, and a sensitivity of $\mathcal{O}(10^{-10})$ should be within reach. A non-negligible background from radiative events, primarily $e^+e^- \rightarrow \tau^+\tau^- \gamma$, is observed for the $\tau^- \rightarrow \mu^-\gamma$ channel. Depending on the ECAL energy resolution, sensitivity at the level or below $10^{-9}$ could be achieved.

\section{Heavy state experimental overview} 
At higher energies, CLFV can be studied both inclusively, e.g. in $pp \rightarrow \mu e + X$, and in a wide variety of heavy state decays. This section will focus on this latter possibility and review the current status and future prospects for searching CLFV in the decays of the $Z$ and Higgs bosons, the top quark and new BSM particles. The high luminosity program at the LHC and the next generation of colliders promise tremendous progress in that area.

\subsection{\texorpdfstring{$Z$}{Z} boson decays}
New physics leading to the lepton flavor violating $Z$ decays can be model-independently parameterized in the context of the Standard Model EFT~\cite{Grzadkowski:2010es}. The leading interactions producing $Z \rightarrow \ell \ell'$ decays at tree level are dimension-six dipole and Higgs operators. These operators also give rise to low-energy LFV transitions, and barring accidental cancellations, LFV muon decays and muon-to-electron conversion largely outperform $Z \rightarrow \mu e$ decays in terms of NP sensitivity. The situation is markedly different for processes involving taus: the $Z \rightarrow \tau e$ and $Z \rightarrow \tau \mu$ have sensitivities comparable to low-energy processes~\cite{Davidson:2012wn}, and provide complementary information.

Searches for flavor violating $Z$ decays into $\mu e$, $\mu \tau$ and $\tau e$ final states have been performed at LEP~\cite{DELPHI97_CLFVZ, OPAL95_CLFVZ}, and more recently by ATLAS~\cite{Aad:2014bca, ATLAS:2020zlz, ATLAS:2021bdj}, setting limits at the level of $0.75-5.0 \times 10^{-6}$. Since these searches are dominated by backgrounds, the limits can be expected to decrease by a factor five or so for 3000 $\ifb$. 

The prospects for LFV $Z$ decays at the FCC-ee were estimated assuming a sample of $3\times 10^{12}$ visible $Z$ decays~\cite{Dam:2018rfz}. A sensitivity on the $Z \rightarrow \mu e$ decays at the level of $10^{-8}$ should be safely within reach, and  one or two order of magnitude improvement could be envisioned with increased PID capabilities. On the $\tau$ side, branching fractions in the vicinity of $10^{-9}$ could be probed for both $Z \rightarrow e \tau$ and $Z \rightarrow \mu \tau$ decays.

\subsection{Higgs boson decays}
Searches for LFV decays of the Higgs boson have been performed by the ATLAS, CMS, and LHCb experiments. Some of these measurements focus on the 125-GeV particle, while other explore lighter or heavier scales. Searches for Higgs boson decays into $e \tau$ or $\mu \tau$ pairs have been performed for different $\tau$ decays channels. Modes with hadronically decaying taus exploit the larger $\tau$ branching fractions, but suffer from significant backgrounds from quark- and gluon-jets. The $H \rightarrow e \mu$ decay is identified as a narrow excess of events over a smooth background, taking advantage from the excellent mass resolution of the lepton pair.

The ATLAS and CMS experiments have established bounds on the $H \rightarrow e \tau$ and $H \rightarrow \mu \tau$ channels at the level of $1.5-4.7 \times 10^{-3}$ at 95\% CL, depending on the final state~\cite{ATLAS:2019pmk,CMS:2021rsq}. These results can be translated into limits on the Yukawa couplings $\sqrt{|Y_{\mu\tau}|^2 + |Y_{\tau\mu}|^2}$ and $\sqrt{|Y_{e\tau}|^2 + |Y_{\tau e}|^2}$, as illustrated in Fig.~\ref{fig:HiggsCMS}. While the constraints on  the $\mu \tau$ channel are already below the theoretical naturalness limit $|Y_{ij}Y_{ji}|=m_im_j/v^2$~\cite{Harnik:2012pb}, the $e \tau$ final state is still an order of magnitude above this threshold. The most stringent limit on the $e \mu$ decay has been established by ATLAS using their full Run-2 dataset, $\mathcal{B} (H \rightarrow e\mu) < 6.1 \times 10^{-5}$~\cite{ATLAS:2019old}. These constraints in the Yukawa couplings are, however, significantly less stringent than those derived from LFV muon transitions.

\begin{figure}[tb]
 \centering
  \includegraphics[width=0.45\textwidth]{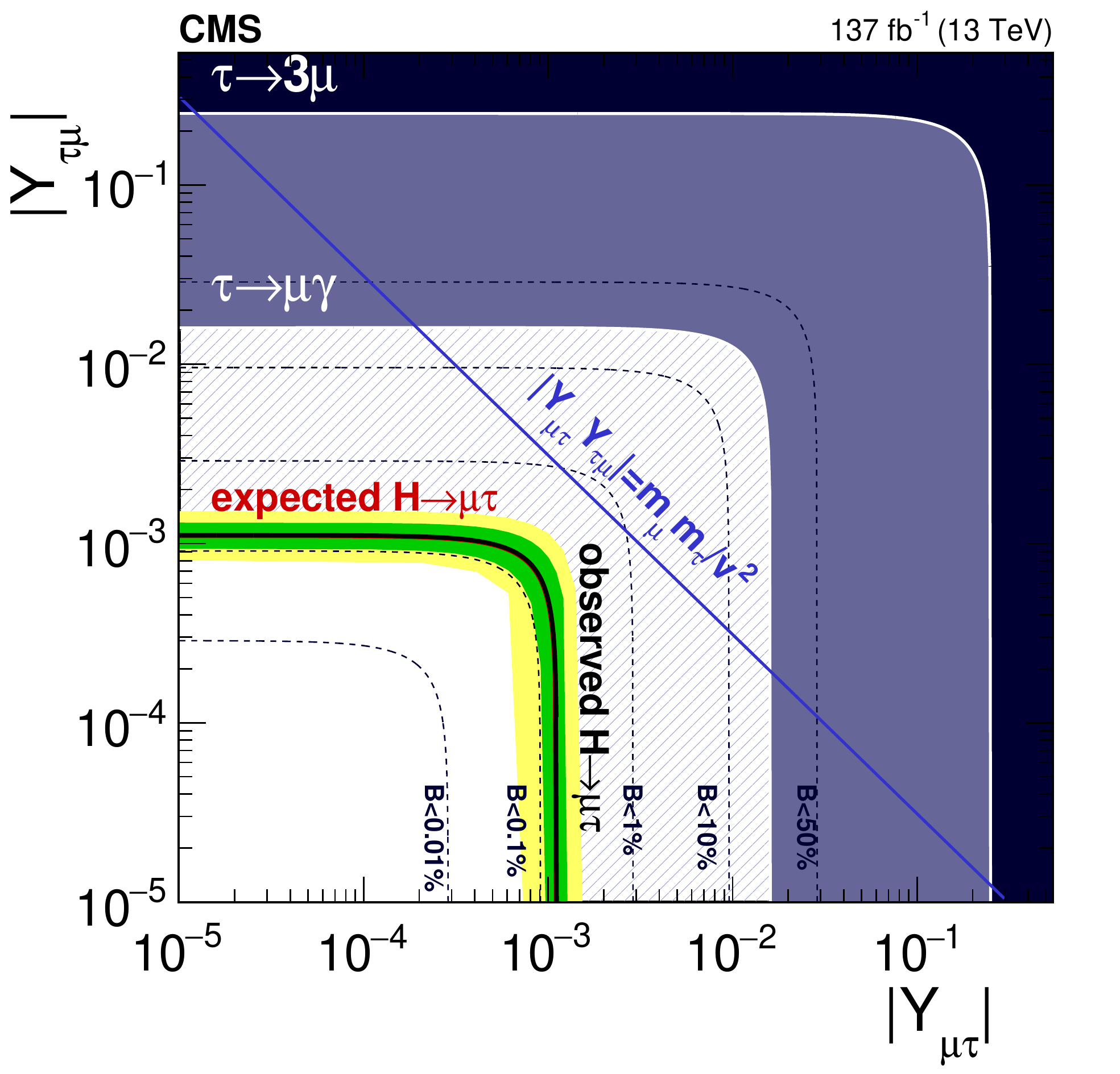}
  \includegraphics[width=0.45\textwidth]{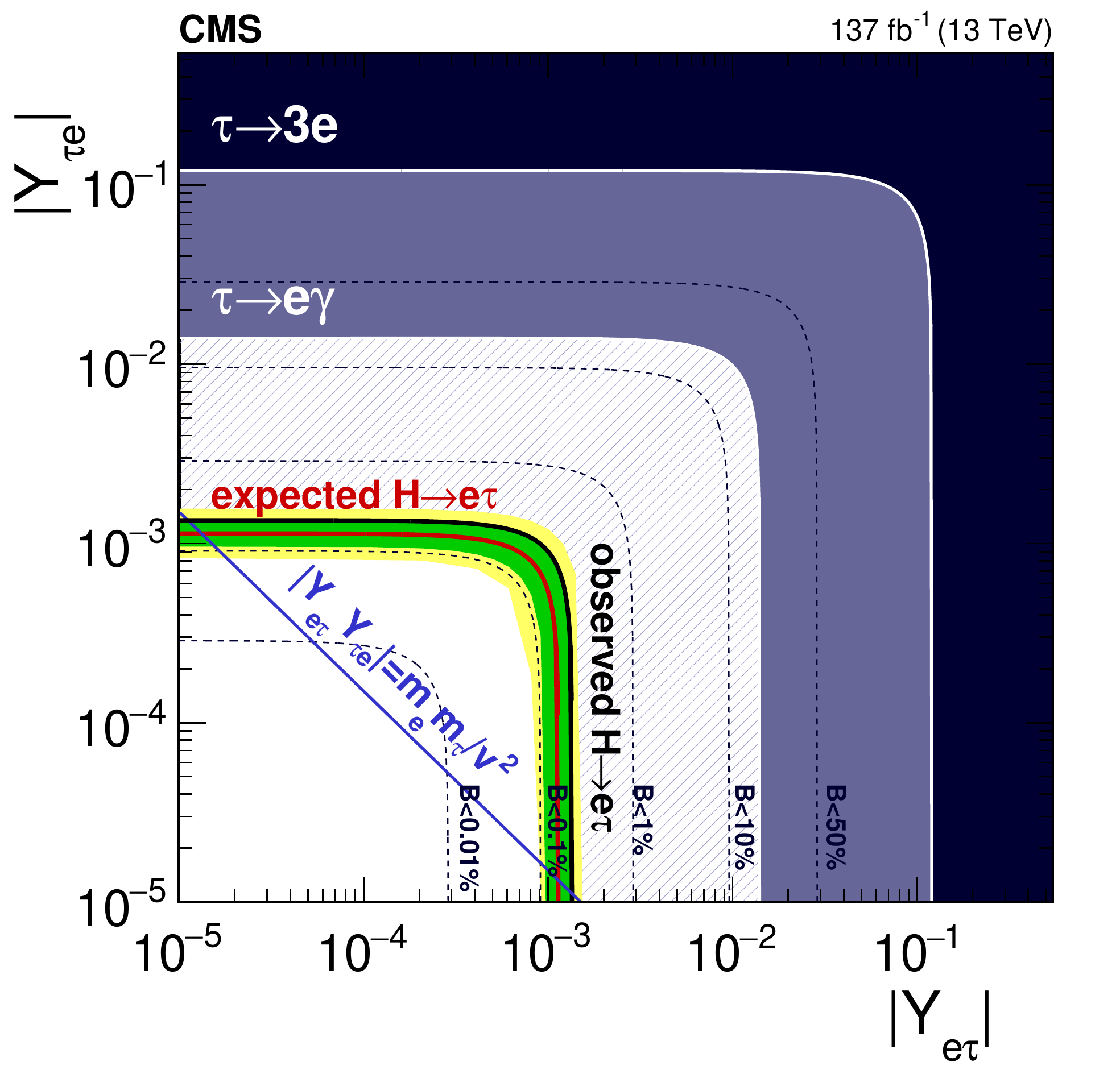}
  \caption{Expected (red line) and observed (black solid line) 95\% CL upper limits on $|Y_{\mu\tau}|$ vs. $|Y_{\tau\mu}|$.  The green and yellow bands indicate the range containing 68\% and 95\% of all observed limit variations from the expected limit. The shaded regions are constraints obtained from null searches for $\tau\to3\mu$ or $\tau\to 3e$ (dark blue) and $\tau\to\mu\gamma$ or $\tau\to e\gamma$ (purple). The blue diagonal line is the theoretical naturalness limit $|Y_{ij}Y_{ji}|=m_im_j/v^2$  (Figure taken from~\cite{CMS:2021rsq}).}
  \label{fig:HiggsCMS}
\end{figure}

Significant improvements are expected at the HL-LHC, and preliminary projections indicate that limit of the order of 0.01–0.05\% could be reached on the $H \rightarrow \tau \mu$ branching fraction~\cite{Cepeda:2019klc}. Sensitivity studies of various $H \rightarrow \ell \ell'$ decays for future $e^+e^-$ colliders were performed assuming the dominant Higgs production mechanism $e^+e^- \rightarrow ZH$ with an event multiplicity of about one million at circular colliders (CEPC and FCC-ee) and half a million at the ILC~\cite{Qin:2017aju}. Upper bounds on the $H \rightarrow e \mu$ and $H \rightarrow \ell \tau$ branching fractions at the level of $1.2 \times 10^{-5}$ ($2.1 \times 10^{-5}$) and $1.5 \times 10^{-4}$ ($2.4 \times 10^{-4}$) were derived for CepC and FCC-ee (ILC), respectively.

\subsection{Top quark decays}
In contrast to the $Z$ and Higgs bosons, top quark LFV decays are necessarily 3-body decays, $t \rightarrow q \ell \ell'$ ($q = u, c$). As these decays must compete with the leading 2-body decay mode $t \rightarrow bW$, their NP reach is slightly lower. However, top decays are highly complementary to the $Z$ and Higgs decays, as they probe qualitatively different types of NP.

More than $10^8$ top quarks were produced during Run II at the LHC. The CMS experiment established bounds on the $t \rightarrow u\mu e$ and $t \rightarrow c\mu e$ branching ratios at the level of $10^{-7}$ and $10^{-6}$, respectively~\cite{CMS:2022ztx}. These results include both searches for top decays as well as searches for single top production in association with $\mu e$. As with other heavy state decays, NP giving rise to LFV top decays can be systematically parameterized by effective dimension-six interactions of the SMEFT. Assuming Wilson coefficients of these operators are of ${\cal O}(1)$, these limits already probe NP above $1 \TeV$~\cite{Grzadkowski:2010es,Davidson:2015zza}.

Some of the operators inducing $t \rightarrow q\ell \ell'$ decays could also be probed by complementary processes. For example, operators containing left-handed $c$ or $u$ quarks can lead to sizable rates in LFV meson decays, such as $K_{L} \rightarrow \mu e$, $B \rightarrow \pi \ell \ell'$ or $B \rightarrow K \ell \ell'$~\cite{Davidson:2015zza}. On the other hand, indirect constraints are very weak for right-handed quark operators, and LFV top decays provide the most sensitive probe.

\subsection{BSM candidates} 
A large variety of BSM scenarios introduce additional neutral scalars $H$ with LFV couplings to the SM charged leptons at the tree-level or 1-loop level~\cite{Hou:1995dg, Dev:2017ftk, Li:2018cod, Arganda:2019gnv}. Given a single LFV coupling $h_{\alpha \beta}$ ($\alpha \neq \beta$), a neutral scalar could be produced in $e^+e^-$ collider via the process $e^+e^- \rightarrow \ell^\pm_\alpha \ell^\mp_\beta H$. If a single Yukawa coupling is non-vanishing, the LFV decays $\ell_\alpha \rightarrow \ell_\beta \gamma$ and $\ell_\alpha \rightarrow 3\ell_\beta$ are forbidden, and only a handful of measurement can constrain $h_{\alpha \beta}$, such as the anomalous magnetic moment of electron~\cite{Mohr:2015ccw} and muon~\cite{Muong-2:2021ojo} or $e^+e^- \rightarrow \ell^+\ell^-$ data~\cite{DELPHI:2005wxt}. Measurements at CEPC and ILC could potentially improve these bounds by up to an order of magnitude~\cite{Dev:2017ftk}.

If two Yukawa couplings are nonzero, and at least one is LFV, the scalar $H$ would induce the LFV process $e^+ e^- \to \ell_\alpha^\pm \ell_\beta^\mp$ at high-energy colliders. While $\mu \to eee$ decays~\cite{SINDRUM:1987nra} have already set stringent constraints on the combination $h_{ee}h_{e\mu}$, colliders could substantially improve the bounds on coupling combinations including taus (e.g. $h_{ee} h_{e\tau}$, $h_{ee} h_{\mu\tau}$ or $h_{e\mu} h_{e\tau}$)~\cite{Dev:2017ftk}.

Heavy or light $Z'$ boson are present in a large variety of NP models with very rich phenomenologies, see e.g.~\cite{Langacker:2000ju, Langacker:2008yv, delAguila:2010mx}. Under the assumption that the $Z$ and $Z'$ LFV couplings to charged leptons are similar, $Z'$ mass up to 5.0 TeV in the $e\mu$ channel, 4.3 TeV in the $e\tau$ channel, and 4.1 TeV in the $\mu\tau$ channel have been excluded~\cite{CMS:2021tau}. The LFV coupling of a light $Z'$ boson can be directly measured at the high-energy hadron and lepton colliders via the process $pp,\, e^+ e^- \to \ell_\alpha^\pm \ell_\beta^\mp Z'$, quite similar to the process discussed for the neutral scalar $H$. FCC-ee could probe couplings down to $10^{-3}$ for light $Z'$ bosons, while masses up to a TeV could be explored at the HL-LHC~\cite{Altmannshofer:2016brv}.

\section{Quark Flavor-changing processes} 

CLFV processes involving quark flavour change, such as $K_L\to e^\pm \mu^\mp$ or $B^+ \to K^+ \tau^\pm e^\mp$, provide a window on NP complementary to quark flavour-diagonal processes (e.g. $\mec$) since quark FCNC are suppressed in the SM. Such decays would be smoking guns for leptoquark models~\cite{Dorsner:2016wpm}, in which generation-diagonal couplings allow tree level decays such as $K \to e^\pm \mu^\mp$ or $K^+\to\pi^+ e^\mp\mu^\pm$. Experimentally, the most recent limits on kaon LFV decays have been set by NA62 with $BR(K^+ \rightarrow \pi^- e^+\mu^+)<4.2\times 10^{-11}$ and $BR(K^+ \rightarrow \pi^+ e^+\mu^-)<6.6\times 10^{-11}$~\cite{NA62:2021zxl}. Slightly better senstivity is achieved by the neutral kaon decays with $BR(K_L\to e^\pm \mu^\mp) < 4.7\times 10^{-12}$~\cite{BNL:1998apv}. The decay modes $B \rightarrow P \ell^\pm \ell'^\mp$ ($P=\pi,K$ and $\ell= \ e,\mu,\tau$) are generally bounded at the level of $10^{-5}$ ($10^{-8}$) for final state including (excluding) a tau lepton.

Leptoquarks are also a popular explanation for Lepton Universality Violation (LUV) observed in B decays~\cite{London:2021lfn}, as they can induce the $(\overline{s}\gamma^\beta P_L  b)(\overline{\mu} \gamma_\beta P_L \mu)$ operator suggested by the data~(see "Weak decays of b and c quarks” Topical Group report~\cite{rapportTG1}). Like other NP models generating LUV, they naturally induce CLFV heavy quark decays, predicting for example $B \rightarrow K \tau \mu$ at the level of $10^{-8}$~\cite{Guadagnoli:2022oxk}. Indeed, a confirmation of NP coupled to muons  would strengthen the physics case for  CLFV searches involving muons (the long-standing anomaly in  the muon magnetic moment~\cite{Aoyama:2020ynm,Muong-2:2021ojo} also suggests BSM coupled to muons), since the NP flavor structure would generically be misaligned with the SM Yukawa couplings~\cite{Calibbi:2021qto, Isidori:2021gqe, Glashow:2014iga}. Anomalies in $B\to \tau \nu X$ decays could already be induced by LFV New Physics~\cite{Ardu:2022pzk} since the neutrino flavor is not observed. The $B$ anomalies make the tantalising suggestion that CLFV involving muons could soon be within  experimental reach.

\section{Conclusion}

Charged lepton flavor violating processes are NP that must occur, and offer a uniquely sensitive gateway to many  scenarios of physics beyond the SM. Together with searches performed at collider, dark matter, and neutrino physics experiments, they provide critical information on the scale and dynamics of flavor generation, and an observation would be unambiguous evidence of NP. 

Existing measurements in the muon sector already probe mass scales at the level of $10^3 - 10^4 \TeV$, and planned experiments will further improve the sensitivity by an order of magnitude. In addition, studies of the muon conversion rate as a function of the target material can provide information about the NP structure. Searches in tau decays can be conducted over many final states, which is promising for identifying the nature of the underlying NP. Reactions involving heavy states ($Z$,$h$,...) and mesons give complementary handles on the structure of physics beyond the SM. 

A global experimental program of CLFV searches is underway in the US, Europe and Asia. Among the most sensitive probes are experiments using high intensity muon beams to search for CLFV transitions, including the coherent neutrinoless conversion of a muon into an electron with Mu2e at FNAL. A staged program of next-generation experiments and facilities has been proposed to exploit the full potential of the PIP-II accelerator. Mu2e-II is a near term evolution of the Mu2e experiment, with an order of magnitude or more improvement in sensitivity to the conversion rate. By leveraging existing infrastructures, Mu2e-II plans to starts construction before the end of the decade. The Advanced Muon Facility is a longer term, more ambitious proposal for a new high-intensity muon complex. This facility would provide the world's most intense positive and negative muon beams, enabling a  suite of experiments with unprecedented sensitivity to probe mass scales in the range $10^4 - 10^5 \TeV$, as well as the possibility to identify the type of operators contributing to New Physics. The development of this complex has also synergies with R\&D for the muon collider and a beam dump dark matter program at FNAL. AMF could be implemented in a phased approach: a conceptual design for a complex with a $\sim 100$ kW proton beam power using PIP-II as a first stage, followed by an upgrade to reach a final $\sim 1$ MW power. A strong R\&D program could make this complex a reality in the next decade.  

The next generation of CLFV experiments and facilities are an essential component of a global program to search for NP. CLFV searches confront the lepton sector in unique way, and may provide the next clues to understanding physics beyond the Standard Model (SM). Strong and continued support of the US community towards current experimental efforts and the development of next-generation experiments and facilities is critical to the realization of the long term physics goals of this program.

\bibliographystyle{Common/JHEP}
\bibliography{Common/references}  

\end{document}